\documentclass[conference]{IEEEtran}
\IEEEoverridecommandlockouts
\usepackage{cite}
\usepackage{amsmath,amssymb,amsfonts}
\usepackage{algorithmic}
\usepackage{graphicx}
\usepackage{multirow}
\usepackage{textcomp}
\usepackage{xcolor}
\usepackage{booktabs}
\usepackage{adjustbox}
\usepackage{array}
\usepackage[caption=false,font=footnotesize]{subfig}
\def\BibTeX{{\rm B\kern-.05em{\sc i\kern-.025em b}\kern-.08em
    T\kern-.1667em\lower.7ex\hbox{E}\kern-.125emX}}
\begin{document}

\title{Data-Driven Multi-Emitter Localization Using Spatially Distributed Power Measurements\\
}

\author{\IEEEauthorblockN{H. Nazim Bicer}
\IEEEauthorblockA{\textit{Department of Electrical Engineering},
\textit{University of Notre Dame, Notre Dame, IN, 46556},
hbicer@nd.edu}
}

\maketitle
\newcommand{\qqi}{\mathbf{q}_i}        
\newcommand{\qhati}{\hat{\mathbf{q}}_i}
\newcommand{\zzi}{\mathbf{z}_i}        
\newcommand{\zhati}{\hat{\mathbf{z}}_i}

\begin{abstract}
With more devices competing for limited spectrum, dynamic spectrum sharing is increasingly vulnerable to interference from unauthorized emitters.
This motivates fast detection and localization of these emitters using low-cost, distributed sensors that do not require precise time synchronization.
This paper presents two convolutional neural network (CNN) approaches for multi-emitter detection and localization from sparsely sampled power maps.
The first method performs single-stage prediction of existence probabilities and positions.
The alternative two-stage method first estimates an occupancy map as an interpretable intermediate representation and then localizes emitters.
A unified training objective combines binary cross entropy with coordinate regression loss and can handle an unknown emitter count.
Small footprint networks, on the order of 70\,k parameters, are trained and evaluated on simulated free-space and urban scenes.
Experiments demonstrate that both approaches localize multiple emitters from sparse measurements across diverse environments, with the logits based two-stage variant remaining competitive, and in some cases superior, under extreme sensor sparsity.
The findings indicate that small CNNs with a unified objective can be deployed for spectrum monitoring and localization.
\end{abstract}
\vspace{-0.35cm}
\section{Introduction}
The wireless spectrum is a vital yet limited resource.
To efficiently utilize the available spectrum, entities like the U.S. Federal Communications Commission (FCC) and the National Telecommunications and Information Administration (NTIA) allocate, assign, and oversee spectrum use.
As the number of devices that harness the wireless spectrum grows rapidly, the dynamic sharing of spectrum in time, frequency, and space has gained importance and has become an active research area~\cite{dynamic_spectrum}.

As requirements become stricter and the complexity of spectrum-sharing algorithms increases, it is essential to have monitoring and mitigation tools to address interference caused by spectral intruders.
These intruders can cause significant financial and safety impacts by obstructing deployed spectrum-sharing systems, especially around time-critical systems like air-traffic control.

A prominent and still publicly unresolved example of this occurred near Dallas--Fort Worth in October 2022~\cite{liu2024localizing}, where global navigation satellite system (GNSS) reception in the airspace near the airport was disrupted for over 20 hours, prompting a temporary runway closure.
In another incident, weather radars operating at 5.60--5.65~GHz experienced interference from non-compliant devices~\cite{weather_radar}.
These incidents often trigger extensive, costly field hunts to locate the source and demonstrate the need for fast detection and precise localization of emitters.

Detecting and localizing harmful interferers is a subject of significant research.
Many prior works assume a known number of sources and a specific propagation model.
Such assumptions limit model generalization across environments with differing channel conditions.
In addition, sensor complexity varies widely, raising practical issues such as time synchronization.
This study considers spatially distributed sensors that use narrowband power measurements to detect and localize emitters which do not require precise time synchronization.
This setup has been considered before in~\cite{deeptxfinder}, where a two-stage process using convolutional neural networks (CNNs) is utilized to detect and localize multiple emitters.
First, a neural network is trained to detect the number of emitters, whose output is used to select the second-stage network that localizes the emitters.
This necessitates training multiple CNNs with matching cardinality in the second stage.
An alternative approach is used in~\cite{deepmtl}, where detection and localization are handled in two stages as image-to-image translation followed by an object-detection task.
The authors use a customized version of the well-known YOLOv3 model~\cite{yolov3} for the second stage as their choice of object detector, with approximately 60 million parameters.

In this study, two data-driven methods are proposed to tackle emitter detection and localization using neural networks with very small footprints, on the order of 70\,k parameters.
In the first method, a single CNN is trained end-to-end for detection and localization.
In the second method, a two-stage pipeline is used. The first network performs occupancy detection, and the following stage localizes emitters using those outputs.
The networks are trained and evaluated on three datasets: (i) free-space, (ii) Chicago, and (iii) Sydney, where the latter two were generated using advanced radio propagation software~\cite{atoll}.

This work presents evidence that accurate multi-emitter localization need not rely on large object-detection networks~\cite{deepmtl} or cardinality specific models~\cite{deeptxfinder}.
CNNs with small footprints trained on power maps from sparse, spatially distributed sensors are found to detect and localize emitters across diverse scenes without prior knowledge of emitter count, power, or a propagation model.
\section{Problem Formulation}
\begin{figure*}[t]
    \centering
    \includegraphics[width=\textwidth]{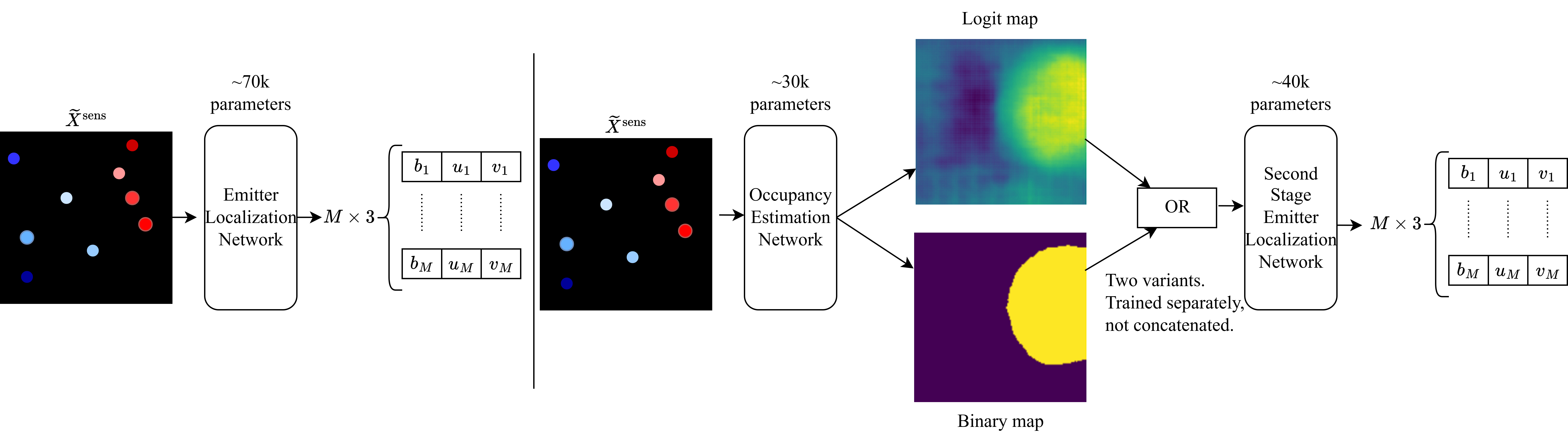}
    \caption{Comparison of single-stage (left) and two-stage (right) emitter localization. $\widetilde{X}^{\mathrm{sens}}$, defined in Eq.\,(4), is used as input for both of the pipelines. The example input image shows measured spectrum power for the sensors as red being high and blue being low. In the two-stage approach, the occupancy estimation network outputs either a logits map or a post-processed binary occupancy map, and only one of these is used per training run by the second-stage localization network.}
    \label{fig:single_two_stage}
\end{figure*}
Let $\mathcal R \subset \mathbb R^2$ be a bounded 2D planar region.
The set of emitter locations is denoted by $\mathcal E=\{\mathbf e_j\}_{j=1}^{K}\subset\mathcal R$, with unknown cardinality $K$.
The cardinality of this set is bounded by a known $M$ ($K\le M$).
Similarly, the set of spatially distributed sensor locations is denoted by $\mathcal S=\{\mathbf r_i\}_{i=1}^{n_{\mathcal S}} \subset \mathcal R$ with cardinality of the set being equal to $n_{\mathcal S}$.
The spectrum power measurement at a position $\mathbf r_i$ is governed by the propagation environment, distance to existing emitters, and the powers of the emitters.
The power measurement at a position $\mathbf r_i$ is denoted by $y_i=p(\mathbf r_i)$.
No measurement noise or modeling error is considered in this study for the received power measurements.

Since processing these continuous real-world positions requires digitization, the planar region $\mathcal R$ is partitioned into $H \times W$ pairwise-disjoint cells $\{G_{u,v}\}_{u=1,\dots,H;\,v=1,\dots,W}$. 
The centroid of $G_{u,v}$ is denoted by $\mathbf c_{u,v}$ and the cell-index map is defined as:
\begin{equation}    
\begin{aligned}
&\mathrm{cell}\colon\ \mathcal R \to \{1,\dots,H\}\times\{1,\dots,W\},\\
&\mathrm{cell}(\mathbf r)\;=\ (u,v)\quad\text{iff } \mathbf r\in G_{u,v}.
\end{aligned}
\end{equation}
The ground-truth power map $P=\{P_{u,v}\}_{u=1,\dots,H;\,v=1,\dots,W}$ is then defined as:
\begin{equation}    
P_{u,v} = \frac{1}{|G_{u,v}|}\int_{G_{u,v}} p(\mathbf r)\,d\mathbf r .
\end{equation}
For each cell, let $I_{u,v}=\{\,i:\mathrm{cell}(\mathbf r_i)=(u,v)\,\}$, and the sensor-sampled map $X^{\mathrm{sens}}\in\mathbb R^{H\times W}$ is defined by:
\begin{equation}
X^{\mathrm{sens}}_{u,v} =
\begin{cases}
\displaystyle \frac{1}{|I_{u,v}|}\sum_{i\in I_{u,v}} y_i, & \text{if } |I_{u,v}|>0,\\[0.9em]
0, & \text{if } |I_{u,v}|=0.
\end{cases}    
\end{equation}
The emitter index set is defined as $\mathcal Z=\{(u_j,v_j)\}_{j=1}^{K}$, which allows for at most one emitter per cell, with $(u_j,v_j)=\mathrm{cell}(\mathbf e_j)$, and the corresponding quantized emitter locations 
$\widehat{\mathcal E}=\{\mathbf c_{u_j,v_j}\}_{j=1}^{K}$.

Given a map representation derived from $X^{\mathrm{sens}}$, the goal of multi-emitter localization is to estimate the unknown index set $\mathcal Z$ and its cardinality $K$.
This equivalently recovers the quantized emitter locations $\widehat{\mathcal E}$.
The result is an approximation of the continuous locations $\mathcal E$ up to grid discretization.

This formulation does not assume knowledge of the number of emitters or their locations. 
The only prior information about the emitters is a known upper bound $M$. 
The available observations are the sensor locations $\mathcal S$ and their power measurements $\{y_i\}_{i=1}^{n_{\mathcal S}}$. 
The grid $\{G_{u,v}\}$ and the bound $M$ are treated as known design choices. 
The map $P$ is a ground-truth quantity used in simulation and is not assumed to be observed.

\section{Proposed Methods}
As shown in Fig.~\ref{fig:single_two_stage}, two data-driven solutions for emitter localization are presented in this section: (i) a single-stage localizer and (ii) a two-stage pipeline.
The rest of this section details the input representation, the model architectures used in each neural network and the training objective utilized during training and training parameters.
\subsection{Input Representation}
As the input representation, a customized version used in \cite{abbas_conf} is utilized.
This representation reduces the dynamic range of input values using a threshold $\tau$ and reduces the complexity of the problem.
The sensor-sampled map $X^{\mathrm{sens}}$ is converted into 
$\widetilde{X}^{\mathrm{sens}}$ by:
\begin{equation}
\widetilde{X}^{\mathrm{sens}}_{u,v}=
\begin{cases}
\displaystyle \frac{1}{10}\Big(\mathrm{dBm}\!\big(X^{\mathrm{sens}}_{u,v}\big)-\tau_{\mathrm{dBm}}\Big),
& \text{if } |I_{u,v}|>0,\\[0.9em]
0, & \text{if } |I_{u,v}|=0,
\end{cases}
\end{equation}
where \(\tau_{\mathrm{dBm}}\in\mathbb R\) is the threshold in dBm and \(\mathrm{dBm}(x)= 10\log_{10}\!\big(x/1\,\mathrm{mW}\big)\).
$\widetilde{X}^{\mathrm{sens}}$ is used as the initial input for both the single-stage and two-stage methods, and it is a single channel image with $H \times W$ shape.

\begin{table*}[t]
\centering
\small
\setlength{\tabcolsep}{3pt}
\renewcommand{\arraystretch}{1.15}
\caption{Single-stage vs two-stage (logit and binary variants) for multi-emitter localization ($M=3$, $n_{\mathcal{S}}=10$). Datasets: Free-space, Chicago,  Sydney. cmr $\uparrow$, rmse $\downarrow$, fa $\downarrow$, and mr $\downarrow$ metrics are reported. The arrows indicate directionality with $\uparrow$ higher is better and $\downarrow$ lower is better.
The best metric for each dataset and metric is shown in \textbf{bold}.}

\begin{tabular}{lcccccccccccc}
\toprule
& \multicolumn{4}{c}{Free-space} & \multicolumn{4}{c}{Chicago} & \multicolumn{4}{c}{Sydney} \\
\cmidrule(lr){2-5} \cmidrule(lr){6-9} \cmidrule(lr){10-13}
Method & CMR$\uparrow$ & RMSE$\downarrow$ & FA$\downarrow$ & MR$\downarrow$ & CMR$\uparrow$ & RMSE$\downarrow$ & FA$\downarrow$ & MR$\downarrow$ & CMR$\uparrow$ & RMSE$\downarrow$ & FA$\downarrow$ & MR$\downarrow$ \\
\midrule
Single-Stage       & 65.38\% & 18.22 & \textbf{7.86\%} & 9.89\% & 63.62\% & 18.98 & 10.12\% & \textbf{8.57\%} & \textbf{64.01\%} & 19.77 & \textbf{10.49\%} & 8.64\% \\
Two-Stage-Logit    & \textbf{68.80\%} & 18.40 & 7.93\% & \textbf{8.25\%} & \textbf{64.26\%} & \textbf{16.23} & \textbf{9.53\%} & 9.06\% & 63.28\% & \textbf{17.63} & 11.43\% & \textbf{8.24\%} \\
Two-Stage-Binary   & 63.92\% & \textbf{17.35} & 8.89\% & 10.29\% & 58.45\% & 16.85 & 11.85\% & 11.43\% & 56.15\% & 19.11 & 12.37\% & 11.97\% \\
\bottomrule
\end{tabular}
\label{tab:single_vs_twostage}
\end{table*}

\subsection{Model Architectures}
All models used in the proposed methods are CNN variants, where an input image is processed using fixed size filters.
The \textit{Occupancy Estimation Network} used in the two-stage method shown on the right side of  Fig.~\ref{fig:single_two_stage} is the same model used in \cite{abbas_conf} with no modifications.
This model uses an encoder-decoder style network to estimate spectrum occupancies and outputs a single channel image with the same dimensions as its input.
The emitter localization networks employed in this work do not require an encoder-decoder structure since the desired output is not an image but rather a set of emitter locations.
Therefore a residual network based CNN is used for emitter localization networks for both the single-stage and the second stage of the two-stage methods.
Both models utilize same filter sizes but the number of channels is increased for the single-stage case to enable fair comparison between single-stage and two-stage networks.
The total parameter budget for each method is approximately 70\;k parameters in total.
\subsection{Loss Function}
The set of emitter index set $\mathcal{Z}$ is an unordered set of locations with cardinality $K$.
When $K<M$, it is necessary to pad the label matrix with $M-K$ emitter labels.
In addition, it is necessary to augment the localization labels with a binary existence flag $b_i \in \{0,1\}, \forall i=1,\dots,M$ to enforce cardinality match at the output of the neural network. 
For the non-existent emitters, the pixel locations are set to $[-1,-1]$.
The resulting set of labels then can be summarized with a matrix of shape $M \times 3$ as shown in Fig.~\ref{fig:single_two_stage}.

The loss function used to train the neural network utilizes two different losses, namely binary cross entropy (BCE) error and mean squared error (MSE).
These loss functions operate on different parts of the labels, with BCE operating on the binary existence flag $b_i$ and MSE operating on the locations.
For a single emitter label $[b_i,u_i,v_i]$, and its corresponding prediction by the network $[\hat{b}_i, \hat{u}_i, \hat{v}_i]$, the loss is calculated as follows:
\begin{equation}    
\qqi=\big(u_i/H,\ v_i/W\big),\qquad
\qhati=\big(\hat u_i/H,\ \hat v_i/W\big),
\end{equation}
\begin{equation}    
\zzi=(b_i,\qqi),\qquad \zhati=(\hat b_i,\qhati).
\end{equation}
\begin{equation}
\ell_i \equiv \ell(\zzi,\zhati)
= \lambda_b\,\mathrm{BCE}(b_i,\hat b_i)
+ \lambda_{xy}\, b_i\, \lVert \qqi-\qhati\rVert_2^{2}.    
\end{equation}
To combat the unordered nature of the output label set, permutation invariant training (PIT) \cite{PIT_paper} is utilized.

\subsection{Dataset Generation and Training}
The datasets used for training, validation, and testing are generated with two simulators. 
The free-space dataset is generated using Friis' transmission equation in a $144~\mathrm{m}\times144~\mathrm{m}$ area at $2.1~\mathrm{GHz}$. 
The Sydney and Chicago datasets are created using an advanced radio-propagation tool~\cite{atoll} with same frequency but in a $864~\mathrm{m}\times864~\mathrm{m}$ area. 
Emitter powers are drawn from $\mathrm{Unif}[20,30]~\mathrm{dBm}$. 
For each neural network, $30720$ samples are used for training, $2048$ for validation, and $2048$ for testing. 
Power maps are resized to single-channel $144\times144$ images setting $H=W=144$. 
Per sample, the number of emitters is drawn from $K\sim\mathrm{Unif}\{1,\dots,M\}$.
$\tau_{\text{dBm}}$ used for input pre-processing is set to $-45$ for free-space and $-95$ for Chicago and Sydney datasets.

All neural networks are trained for $100$ epochs using the Adam optimizer with batch size $64$ and a learning rate of $5\times10^{-4}$. 
For emitter-localization networks, the loss weights are assigned as $\lambda_b=0.1$ and $\lambda_{xy}=0.9$. 
For the occupancy estimation network in the proposed two-stage method, BCE loss is used as in~\cite{abbas_conf}.
After training, the weights are frozen. 
Pixel-wise accuracies for the occupancy networks at $M=3$, $n_{\mathcal S}=10$ are $95\%$, $85\%$, and $87\%$ for free-space, Chicago, and Sydney, respectively. 
The second-stage localization networks are then trained on the predicted outputs of the fixed occupancy networks. 
The model checkpoint with the best validation performance is used for final testing for each network.

\subsection{Evaluation Metrics}
Evaluation is based on cardinality match rate (CMR), root-mean-square error (RMSE, in pixels), false-alarm rate (FA), and miss rate (MR), similar to prior work \cite{deeptxfinder,deepmtl}.
CMR is the fraction of test samples for which the predicted emitter count is equal to the ground truth emitter count.
The emitter is counted as predicted if its existence probability $\hat b_i$ is larger than $\gamma=0.5$.
FA is computed as the total number of extra predicted emitters summed over all samples divided by the total number of ground truth emitters.
MR is the total number of missed emitters divided by the same denominator.
RMSE is computed only for the samples where the cardinality match is achieved.
For such samples, the predicted and true emitter locations are paired with respect to minimum matching error and then averaged over all such samples.

\section{Results and Discussion}
The performance of two-stage method is compared with the performance of single-stage network in an extremely sparse sensor count scenario with $n_\mathcal{S}=10$ and $M=3$ in three different radio propagation environments. 
The results are presented in Table~\ref{tab:single_vs_twostage}.
It is noteworthy that the two-stage design with logit maps as the intermediate step results in comparable and sometimes better performance than the single-stage method with fewer parameters utilized for localization task.
The binary variant, although resulting in lower RMSE for free-space, consistently fails to achieve  comparable results for other metrics compared to other two methods.
This suggests that using hard thresholding in the intermediate step eliminates useful information that could be used by the following stage.
Using occupancy logits as the usable intermediate step provides a good proxy for localization.
This might be attributed to logit map acting as a regularizer that preserves the necessary information.

\begin{figure}[t]
    \centering
    \includegraphics[width=\columnwidth]{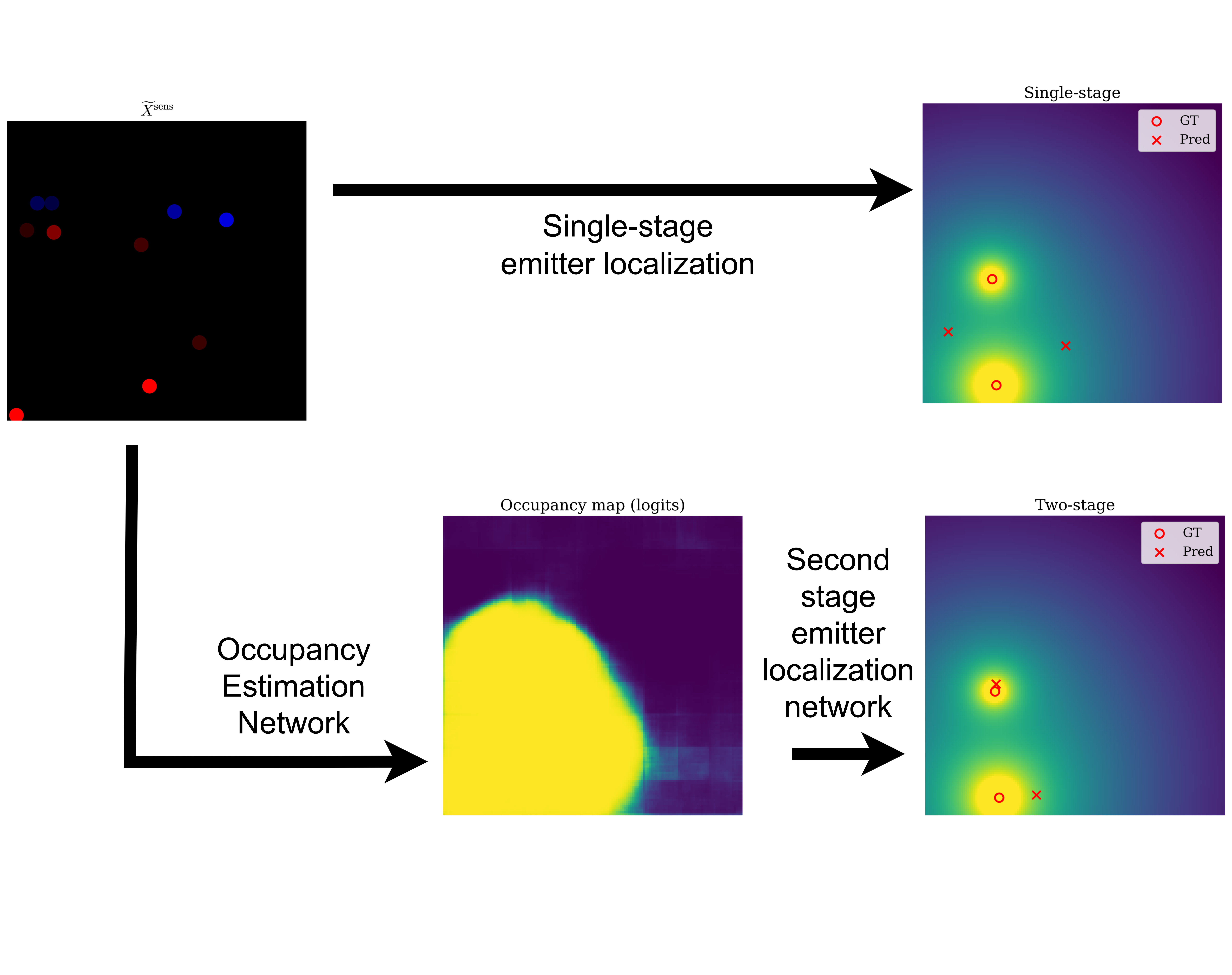}
    \caption{An inference example for single-stage and two-stage with logits variant using same initial input $\widetilde{X}^{\mathrm{sens}}$.
    The top branch shows the single-stage network directly predicting emitter locations from $\widetilde{X}^{\mathrm{sens}}$, 
    while the bottom branch first estimates the occupancy map, followed by a second-stage localization network. The sample presented shows that two-stage method is able to be more accurate compared to single-stage method.}
    \label{fig:single_two_stage_inference_visual}
\end{figure}

\begin{figure}[t]
  \centering
  \subfloat[]{\includegraphics[width=0.485\columnwidth]{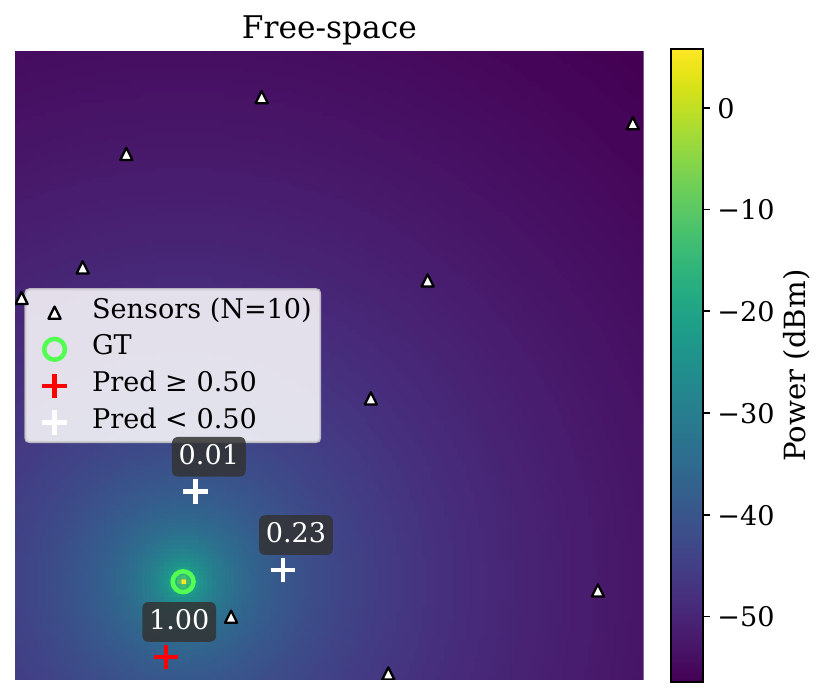}}\hfill
  \subfloat[]{\includegraphics[width=0.485\columnwidth]{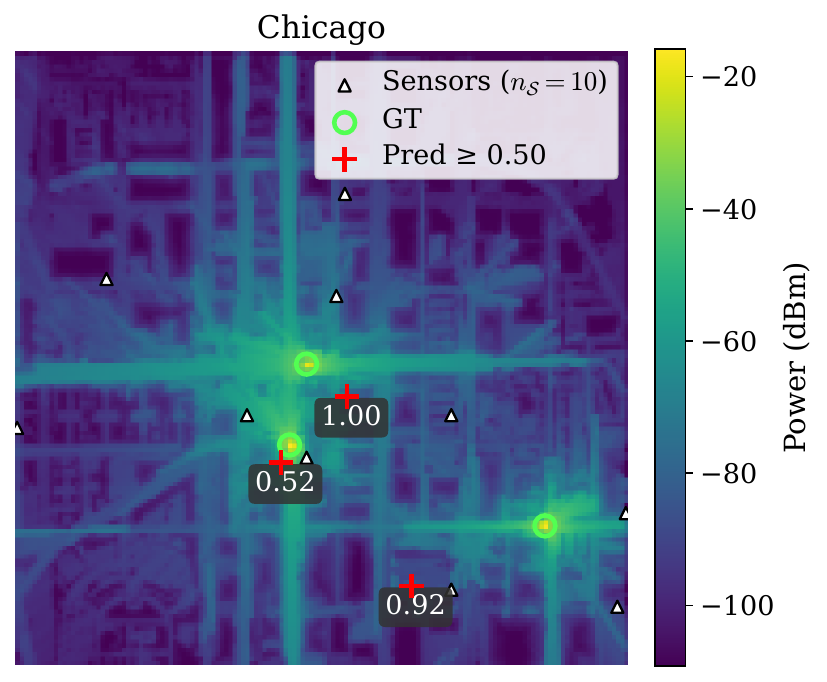}}
  
  \makebox[\columnwidth][c]{%
    \subfloat[]{\includegraphics[width=0.485\columnwidth]{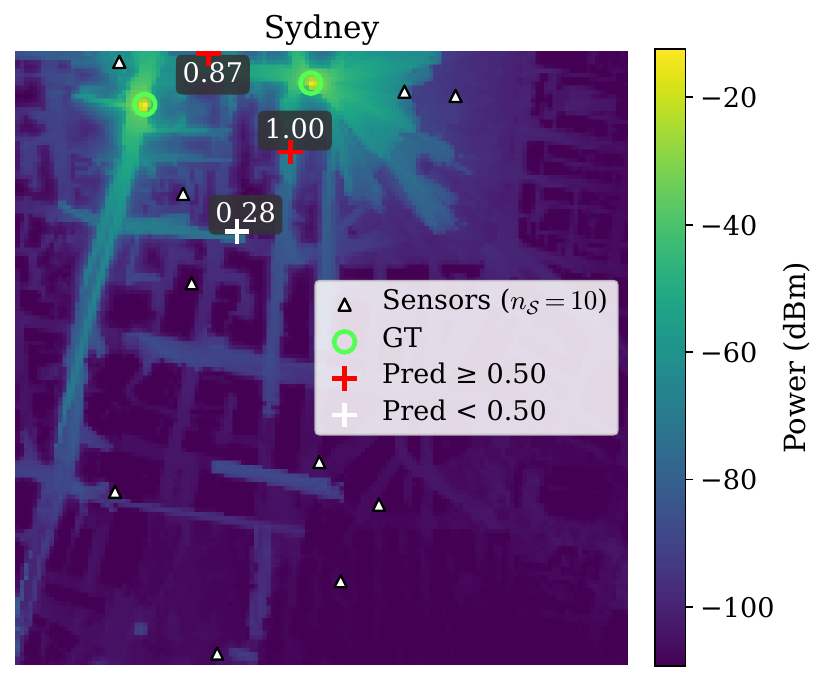}}%
  }
\caption{Two-stage model prediction visualizations over power maps. Models use $M=3$ output slots and is evaluated with $n_{\mathcal{S}}=10$ sensors. Green rings mark ground truth emitters. Red “+” indicate predictions with existence probability $\ge 0.50$ (white “+” if $<0.50$). Numbers annotate predicted existence probabilities. (a) Free-space: A single existing emitter is localized accurately and the remaining slots are low-confidence. (b) Chicago: Three emitters are present and all are predicted to be existing, achieving cardinality match. However, lower-right emitter is localized poorly. (c) Sydney: Two emitters are predicted although localized with low accuracy. The prediction suppresses the third slot.}
\label{fig:visual-preds}
\end{figure}

Fig.~\ref{fig:single_two_stage_inference_visual} shows an example inference sample with two-stage method outperforming the single-stage method and Fig.~\ref{fig:visual-preds} demonstrates examples of the two-stage method across three datasets. 

\section{Conclusion}
This work tackles multi-emitter localization from narrowband power maps using two CNN based methods.
One model performs direct single-stage prediction, and the other uses a two-stage pipeline with an intermediate occupancy step. A permutation-invariant loss merges existence classification with coordinate regression and handles unknown cardinality without external assignment or cardinality-specific models.
Across free-space and urban scenes with sparse sensors, both approaches localize reliably, with the logits based variant performing slightly better in some settings.
These properties support scalable, low-cost deployments across large sensing grids.
\bibliographystyle{IEEEtran}
\bibliography{refs}

@ARTICLE{dynamic_spectrum,
  author={Bhattarai, S. and Park, Jung-Min Jerry and Gao, Bo and Bian, Kaigui and Lehr, William},
  journal={IEEE Transactions on Cognitive Communications and Networking}, 
  title={An Overview of Dynamic Spectrum Sharing: Ongoing Initiatives, Challenges, and a Roadmap for Future Research}, 
  year={2016},
  volume={2},
  number={2},
  pages={110-128},
  doi={10.1109/TCCN.2016.2592921}}

@inproceedings{liu2024localizing,
  author    = {Liu, Zixi and Lo, Sherman and Blanch, Juan and Walter, Todd},
  title     = {Localizing the October 2022 Texas Jamming Incident Using {ADS-B} Data with an Improvement in Model Confidence},
  booktitle = {Proceedings of the 2024 International Technical Meeting of The Institute of Navigation},
  year      = {2024},
  pages     = {524--531},
  address   = {Long Beach, California},
  month     = jan,
  publisher = {Institute of Navigation},
  doi       = {10.33012/2024.19558}
}

@techreport{weather_radar,
  author      = {Carroll, John E. and Sanders, Frank H. and Sole, Robert L. and Sanders, Geoffrey A.},
  title       = {Case Study: Investigation of Interference into 5 {GHz} Weather Radars from Unlicensed National Information Infrastructure Devices, Part I},
  institution = {National Telecommunications and Information Administration (NTIA), Institute for Telecommunication Sciences (ITS)},
  type        = {NTIA Technical Report},
  number      = {TR-11-473},
  address     = {Boulder, CO},
  month       = nov,
  year        = {2010},
  url         = {https://its.ntia.gov/publications/download/11-473.pdf}
}

@INPROCEEDINGS{deeptxfinder,
  author={Zubow, Anatolij and Bayhan, Suzan and Gawłowicz, Piotr and Dressler, Falko},
  booktitle={2020 29th International Conference on Computer Communications and Networks (ICCCN)}, 
  title={DeepTxFinder: Multiple Transmitter Localization by Deep Learning in Crowdsourced Spectrum Sensing}, 
  year={2020},
  volume={},
  number={},
  pages={1-8},
  doi={10.1109/ICCCN49398.2020.9209727}}

@INPROCEEDINGS{deepmtl,
  author={Zhan, Caitao and Ghaderibaneh, Mohammad and Sahu, Pranjal and Gupta, Himanshu},
  booktitle={2021 IEEE 22nd International Symposium on a World of Wireless, Mobile and Multimedia Networks (WoWMoM)}, 
  title={DeepMTL: Deep Learning Based Multiple Transmitter Localization}, 
  year={2021},
  volume={},
  number={},
  pages={41-50},
  doi={10.1109/WoWMoM51794.2021.00017}}

@article{yolov3,
  title   = {YOLOv3: An Incremental Improvement},
  author  = {Redmon, Joseph and Farhadi, Ali},
  journal = {arXiv preprint arXiv:1804.02767},
  year    = {2018}
}

@misc{atoll,
  title        = {Atoll Radio Planning Software},
  howpublished = {[Online]. Available: \url{http://www.forsk.com/atoll}},
  note         = {Accessed: 2025-08-10}
}

@INPROCEEDINGS{abbas_conf,
  author={Termos, Abbas and Hochwald, Bertrand},
  booktitle={2021 IEEE International Symposium on Dynamic Spectrum Access Networks (DySPAN)}, 
  title={Robust Neural Network-Based Spectrum Occupancy Mapping}, 
  year={2021},
  volume={},
  number={},
  pages={296-301},
  doi={10.1109/DySPAN53946.2021.9677439}}

@INPROCEEDINGS{PIT_paper,
  author={Yu, Dong and Kolbæk, Morten and Tan, Zheng-Hua and Jensen, Jesper},
  booktitle={2017 IEEE International Conference on Acoustics, Speech and Signal Processing (ICASSP)}, 
  title={Permutation invariant training of deep models for speaker-independent multi-talker speech separation}, 
  year={2017},
  volume={},
  number={},
  pages={241-245},
  doi={10.1109/ICASSP.2017.7952154}}
\end{document}